\documentclass{PoS}

\pdfoutput=1

\title{Combining NNPDF3.0 and NNPDF2.3QED through the APFEL evolution code}

\ShortTitle{Combining NNPDF3.0 and NNPDF2.3QED through the APFEL evolution code}

\author{\speaker{Valerio Bertone}\,\thanks{OUTP-16-16P}\\
        Rudolf Peierls Center for Theoretical Physics, 1 Keble Road,\\
        University of Oxford, OX1 3NP Oxford, United Kingdom\\
        E-mail: \email{valerio.bertone@physics.ox.ac.uk}}

\author{Stefano Carrazza\thanks{CERN-TH-2016-144}\\
        Theoretical Physics Department, CERN, Geneva Switzerland\\
        E-mail: \email{stefano.carrazza@cern.ch}}

      \abstract{We present sets of parton distribution functions
        (PDFs), based on the NNPDF3.0 family, which include the photon
        PDF from the NNPDF2.3QED sets, and leading-order QED
        contributions to the DGLAP evolution as implemented in the
        public code {\tt APFEL}. The aim is to combine our
        state-of-the-art determination of quark and gluon PDFs with
        the so far only direct determination of the photon PDF from
        LHC data. In addition, the use of {\tt APFEL} allowed us to
        employ a solution of the DGLAP equation that, differently from
        that used for the NNPDF2.3QED sets, includes QED corrections
        in a more accurate way. We briefly discuss how these sets are
        constructed and investigate the effect of the inclusion of the
        QED corrections on PDFs and parton luminosities. Finally, we
        compare the resulting sets, which we dubbed NNPDF3.0QED, to
        the older NNPDF2.3QED sets and to all presently available PDF
        sets that include QED corrections, namely CT14QED and
        MRST2004QED.}

\FullConference{DIS 2016
		11-15 April 2016\\
		Hamburg, Germany}

\begin{document}

\paragraph{Introduction}\label{sec:intro}

It has been shown in many contexts that the inclusion of electroweak
(EW) corrections is a crucial requirement for precision phenomenology
at the LHC and at the Future Circular Collider (FCC) (see $e.g.$
Refs.~\cite{Pagani:2016caq,Bertone:2015lqa}). As a consequence, a huge
effort is being put in the computation of EW corrections to hard
matrix elements. However, a consistent inclusion of such corrections
at the level of physical predictions requires the use of PDF sets
which incorporate QED effects. This in particular implies the
inclusion of QED corrections to the perturbative DGLAP evolution, and
thus the presence of a photon PDF.

Presently, several collaborations provide sets with QED
corrections. Historically, the very first PDF set that implemented QED
corrections was the pioneering MRST2004QED set~\cite{Martin:2004dh}
whose photon content was determined by assuming that the respective
distribution at the initial scale is obtained by one-photon collinear
emission off valence quarks, such that the only parameters of this
model are the masses of the \textit{up} and \textit{down} quarks.
The MRST2004QED set provides two estimates of the photon PDF based on
two different assumptions for the \textit{up} and \textit{down} quark
masses (``current'' and ``constituent'') but no experimental
uncertainty on the respective distributions is given.

The next PDF set in chronological order to contain a photon PDF was
the NNPDF2.3QED set~\cite{Ball:2013hta}. Differently from MRST2004QED,
the NNPDF2.3QED photon PDF was parametrized on the same footing as all
other partons and no theoretical assumption on its functional form was
made. The photon PDF was determined by a direct fit to DIS data and by
reweighting on LHC Drell-Yan data. As a consequence, the resulting
photon PDF was delivered with an experimental uncertainty. It is worth
noticing that the QED corrections to the DGLAP evolution in
NNPDF2.3QED were implemented in a different way with respect to the
MRST2004QED set. The difference stems from the fact that in the former
the QCD and the QED factorization scales are taken to be independent
and the evolution with respect to each scale is done successively,
while in the latter these scales coincide and the respective
evolutions are done simultaneously.

More recently the CT14QED set~\cite{Schmidt:2015zda} was made
public. In this set the photon PDF is determined employing the same
theoretical ansatz of the MRST2004QED set. Again, no real fit of the
photon PDF is done but, differently from the MRST2004QED set, the
ansatz for the photon PDF has one single parameter identified with the
momentum fraction carried by the photon at the initial scale
$p_0^{\gamma}$. This parameter is estimated by comparison with DIS
data for isolated photon production from the ZEUS
experiment~\cite{Chekanov:2009dq}, resulting in the constraint $0\% <
p_0^{\gamma} < 0.14\%$ at the 90\% confidence level (CL).


\paragraph{The NNPDF3.0QED set}\label{sec:nnpdf30qed}

The purpose of this contribution is to document a new set of PDFs with
QED corrections. This new set is based on the recent NNPDF3.0 global
analysis~\cite{Ball:2014uwa} and incorporates the photon PDF from the
NNPDF2.3QED analysis, employing the {\tt APFEL} code for the
QED-corrected DGLAP evolution: we dubbed it NNPDF3.0QED.

As already pointed out, the evolution of the NNPDF2.3QED sets is such
that the subtraction of the QCD and QED collinear divergences is done
separately. This implies the introduction of two different
factorization scales, $\mu_{F,\rm QCD}$ and $\mu_{F,\rm QED}$, and the
DGLAP evolution with respect to each of them is done sequentially and
independently. This approach, as compared to the more common procedure
adopted in the MRST2014QED and CT14QED sets in which QCD and QED
factorization scales are identified, leads to a suppression of the
photon PDF at large scales and small values of the Bjorken
$x$~\cite{Schmidt:2015zda,Pagani:2016caq}. In order to construct the
NNPDF3.0QED sets we have dropped the distinction between QCD and QED
factorization scales and, in line with the other collaborations, we
have adopted the so-called QCD$\otimes$QED \textit{unified} solution
of the DGLAP equation as implemented in the public code {\tt
  APFEL}~\cite{Bertone:2013vaa} in which QCD and QED evolutions are
done simultaneously ensuring a better accuracy.

The NNPDF3.0QED sets are constructed by combining the QCD partons,
$i.e.$ gluon and quark PDFs, from the NNPDF3.0 sets with the photon
PDF from the NNPDF2.3QED sets. On the one hand, the NNPDF3.0 sets provide
a state-of-the-art determinations of quark and gluon PDFs. On the
other hand, the photon PDF extracted in the NNPDF2.3QED analysis is
currently the only determination entirely based on fits to DIS and LHC
data without the assumption of any model. Although the combination of
PDFs extracted from different analyses introduces a potential
inconsistency, the low level of correlation between the photon and the
other PDFs guarantees that no large inaccuracy is introduced. In
addition, as we will show in the following, the smallness of the
photon PDF ensures that the momentum sum rule (MSR) is not
significantly violated.

The combination of the NNPDF3.0 QCD partons with the NNPDF2.3QED photon is
done at the scale $Q = \sqrt{2}\simeq 1.414$ GeV and the resulting
sets of PDFs are then evolved using the QCD$\otimes$QED unified
solution provided by {\tt APFEL}. This procedure is applied both at
NLO and NNLO in QCD, while the QED corrections are always accounted to
LO. It is worth mentioning that all NNPDF sets implement the so-called
\textit{truncated} solution of the DGLAP equation~\cite{Vogt:2004ns}
in which the evolution operator is expanded in powers of the strong
coupling $\alpha_s$ and truncated to the required order. This
particular solution is implemented in {\tt APFEL} and it has been used
to produce the NNPDF3.0QED sets\footnote{In the NNPDF sets also the
  evolution of the strong coupling $\alpha_s$ is implemented by means
  of an iterative analytic solution of the RG equation based on a
  perturbative expansion in powers of $\alpha_s$ (see $e.g.$
  Ref.~\cite{DelDebbio:2007ee}).}.

\begin{figure}
  \begin{centering}
    \includegraphics[scale=0.35]{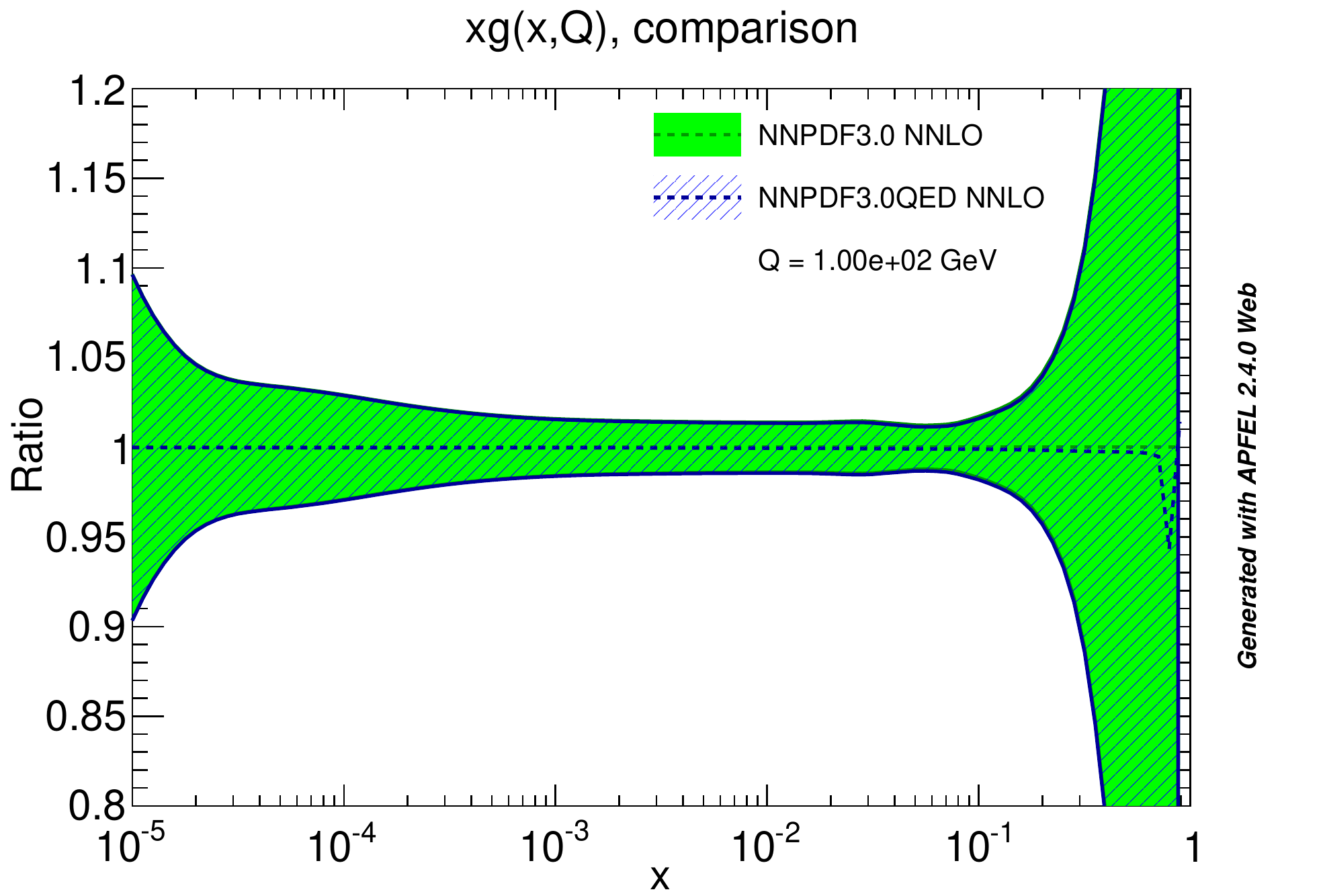}\includegraphics[scale=0.35]{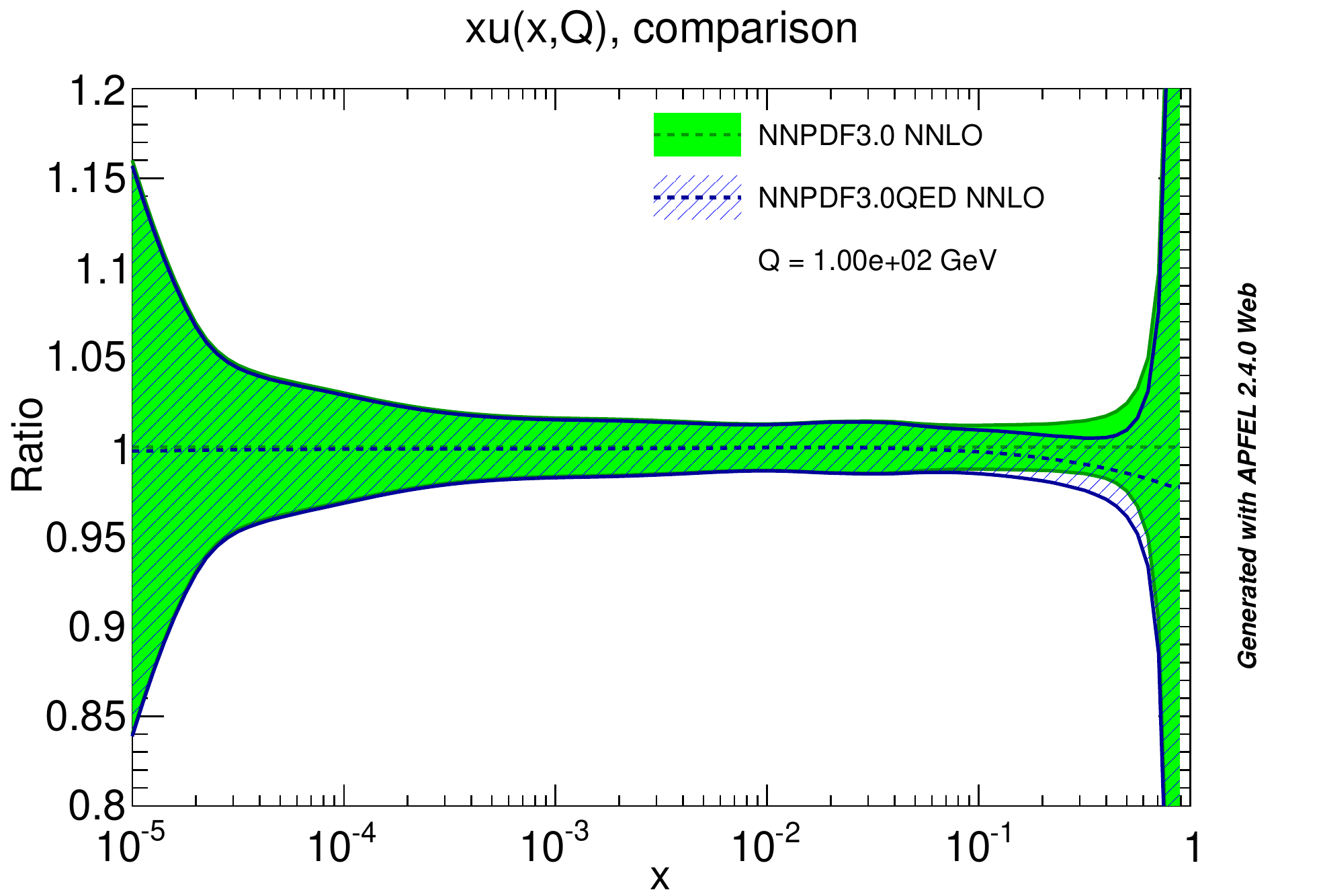}
    \includegraphics[scale=0.35]{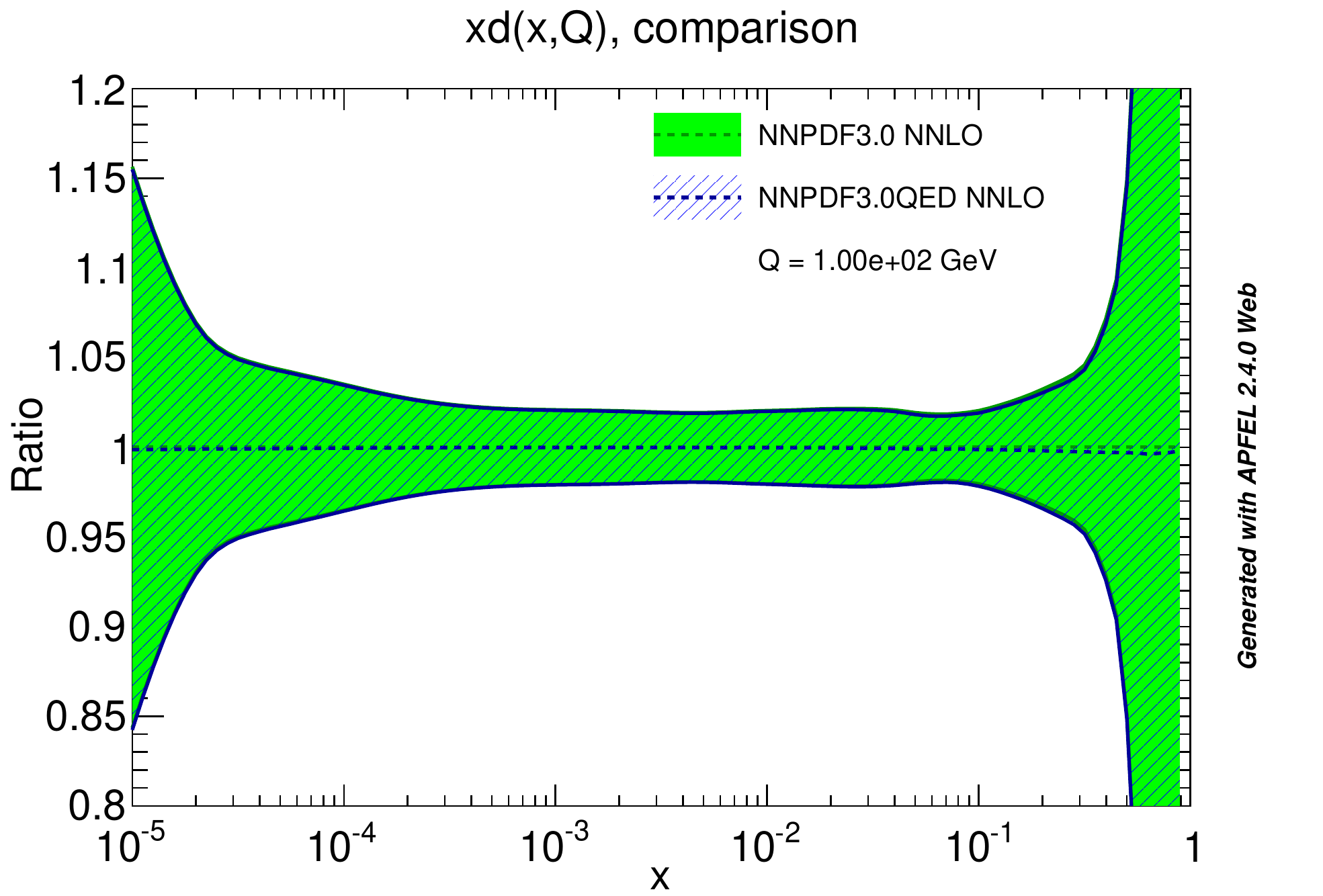}\includegraphics[scale=0.35]{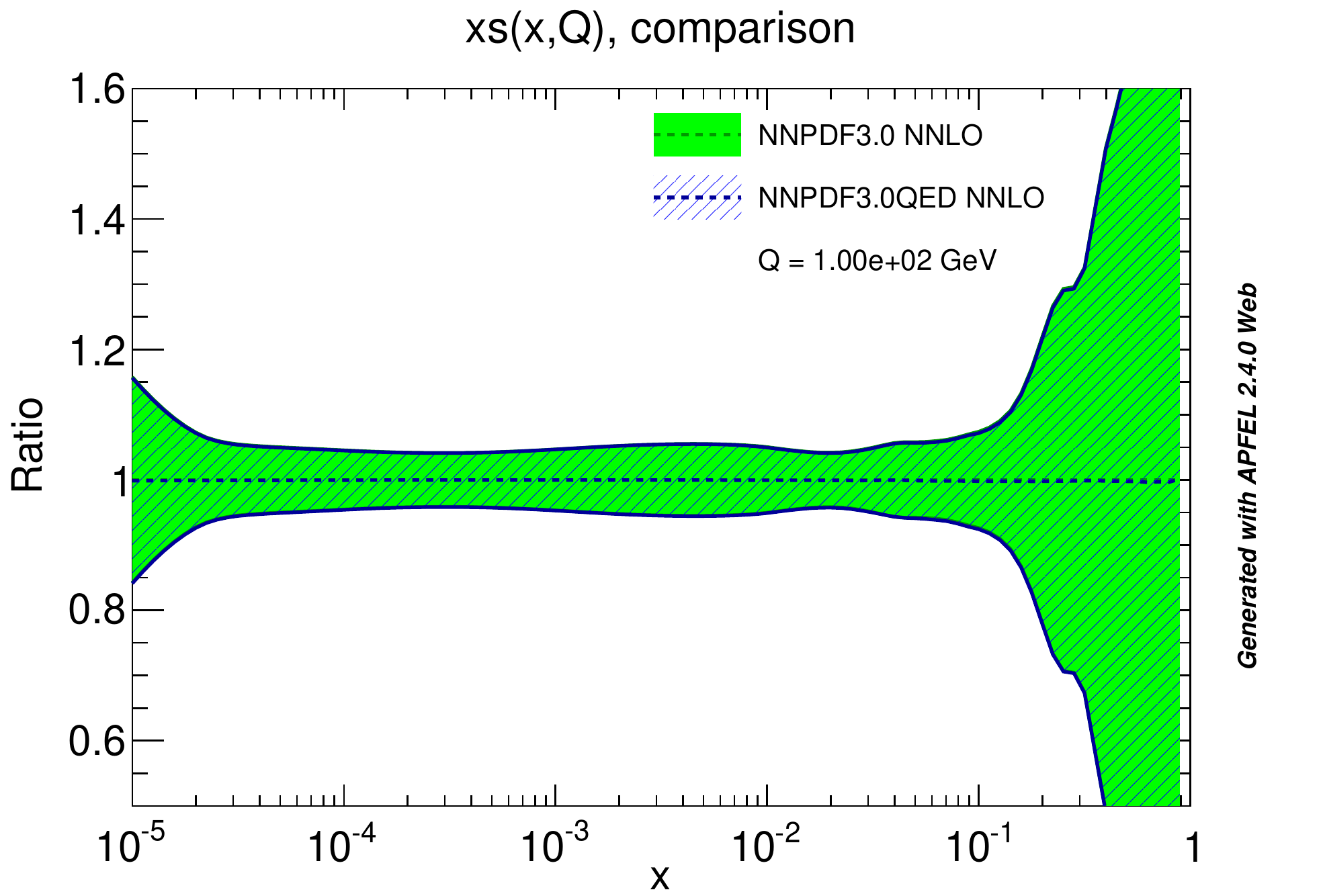}     
    \par\end{centering}
  \caption{\label{fig:pdfcomparison}\small Comparison between NNPDF3.0
    NNLO and NNPDF3.0QED NNLO for the $g$, $u$, $d$, and $s$
    distributions at $Q=100$ GeV. Plots generated with {\tt APFEL
      Web}~\cite{Carrazza:2014gfa}.}
\end{figure}

In Fig.~\ref{fig:pdfcomparison} we compare the gluon, up, down, and
strange distributions from the NNPDF3.0QED set at NNLO to the
respective distributions from the NNPDF3.0 set at $Q = 100$ GeV. The
plots, presented as ratios to NNPDF3.0, show that the effect of the
QED corrections to the DGLAP evolution on the QCD partons is very
small. It is interesting to observe that, although the deviations are
very limited everywhere, the up PDF is the relatively most affected of
the quark distributions. This is consistent with the fact that the QED
coupling is proportional to the squared charge $e_q^2$ of the quark
and thus the up-type quarks are more affected than the down-type
ones. Similar results are obtained at NLO.

\begin{figure}
 \begin{centering}
   \includegraphics[scale=0.35]{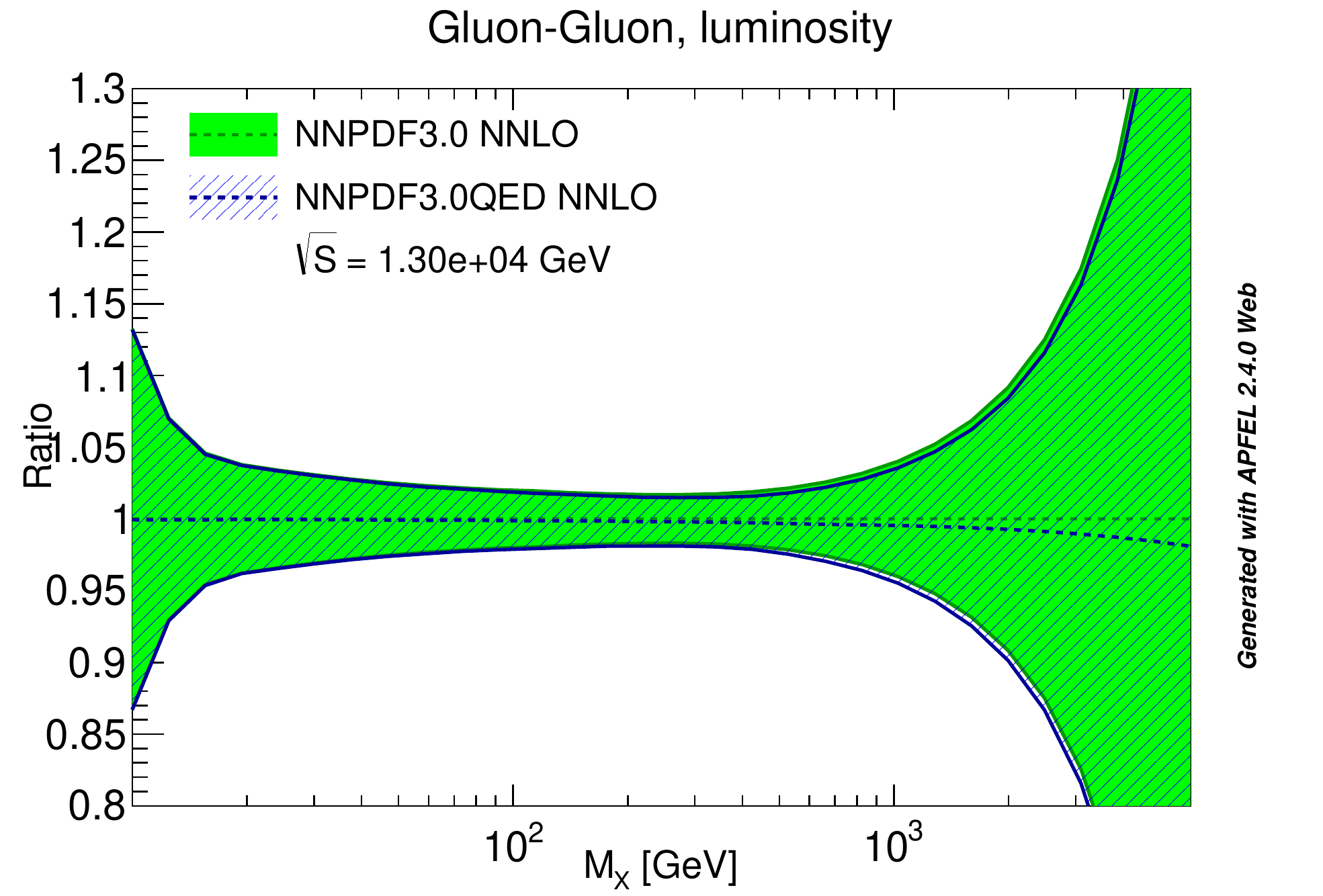}\includegraphics[scale=0.35]{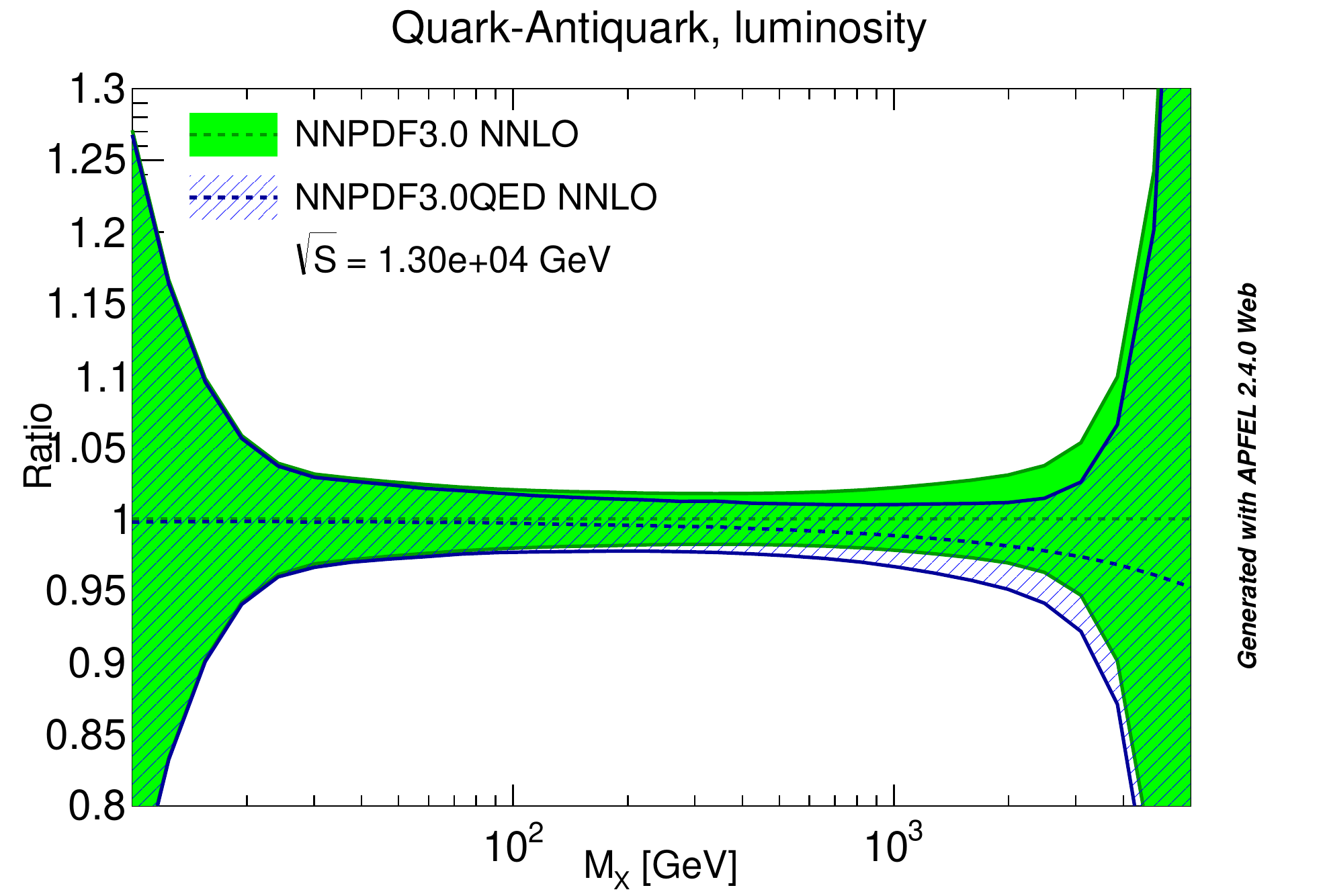}
   \par\end{centering}
 \caption{\label{fig:luminosities}\small Comparison between NNPDF3.0
   NNLO and NNPDF3.0QED NNLO for the $gg$ and $q\bar{q}$ parton
   luminosities at $\sqrt{s} = 13$ TeV as functions of the final state
   in variant mass $M_x$.}
\end{figure}

Fig.~\ref{fig:luminosities} shows the comparison between NNPDF3.0 and
NNPDF3.0QED at NNLO for the gluon-gluon and the quark-antiquark parton
luminosities at the center of mass energy $\sqrt{s} = 13$ TeV as
functions of the final state in variant mass $M_X$. The plots are
presented as ratios to NNPDF3.0. As expected, the introduction of the
QED corrections has a relatively small impact also on the parton
luminosities. In par\-ti\-cu\-lar, the gluon-gluon luminosity is
essentially unaffected, while the quark-antiquark luminosity from
NNPDF3.0QED presents a suppression with respect to that of the
NNPDF3.0 set for large values of $M_X$. This is the consequence of the
change in the quark PDFs for large values of $x$ shown in
Fig.~\ref{fig:pdfcomparison}. The effect is however well within
uncertainties.

We now turn to consider the photon PDF of the NNPDF3.0QED sets. As a
first step we check that the inclusion of the photon PDF to the
NNPDF3.0 sets does not lead to any significant violation of the
MSR. For the NNLO sets we find:
\begin{equation}\label{eq:msr}
\small
\begin{array}{ll}
\mbox{NNPDF3.0 NNLO:} & \displaystyle \int_0^1 dx
                        \,x\left[g(x,Q)+\sum_{i=1}^6 q_i(x,Q)+\overline{q}_i(x,Q)\right]
                        = 0.999\pm 0.005\,,\\
\\
\mbox{NNPDF3.0QED NNLO:} & \displaystyle \int_0^1 dx
                        \,x\left[\gamma(x,Q)+g(x,Q)+\sum_{i=1}^6 q_i(x,Q)+\overline{q}_i(x,Q)\right]
                        = 1.002\pm 0.008\,,
\end{array}
\end{equation}
and thus the MSR is conserved within uncertainties\footnote{Note that
  the DGLAP evolution preserves the total momentum fractions and thus
  the MSR is independent of the scale $Q$.}. Similar results are found
at NLO. This indicates that the momentum fraction carried by the
photon is very small, it is however interesting to quantify it. In
Tab.~\ref{tab:momfractions} we report the percent photon momentum
fraction defined as:
\begin{equation}
p^{\gamma}(Q) = \int_0^1dx\,x\gamma(x,Q)\,,
\end{equation}
at NLO and NNLO for three different scales.  As expected, the fraction
of the momentum carried by the photon grows slowly with the scale but
the associated uncertainties are typically large. In particular, at
low scales the photon momentum fraction is nearly compatible with
zero.
\begin{table}
\begin{centering}
\begin{tabular}{|l|c|c|c|}
\hline 
PDF Set & $p^{\gamma}(Q=\sqrt{2}\mbox{ GeV})$ & $p^{\gamma}(Q=10^2\mbox{ GeV})$ & $p^{\gamma}(Q=10^3\mbox{ GeV})$ \tabularnewline 
\hline 
\hline 
NNPDF3.0QED NLO & $0.4  \pm   0.4$ \% & $0.6  \pm   0.4$ \% & $0.7 \pm
                                                              0.4$ \% \tabularnewline 
\hline 
NNPDF3.0QED NNLO & $0.4  \pm   0.3$ \% & $0.6 \pm 0.3$ \%& $0.7 \pm
                                                           0.3$ \% \tabularnewline 
\hline 
\end{tabular}
\par\end{centering}
\caption{\label{tab:momfractions}\small Photon momentum fractions at $Q
  =\sqrt{2},10^2,10^3$ GeV for NNPDF3.0QED NLO and NNLO. Uncertainties
  are given at 68\% CL symmetric around the central value.}
\end{table}

\begin{figure}
  \begin{centering}
    \includegraphics[scale=0.55]{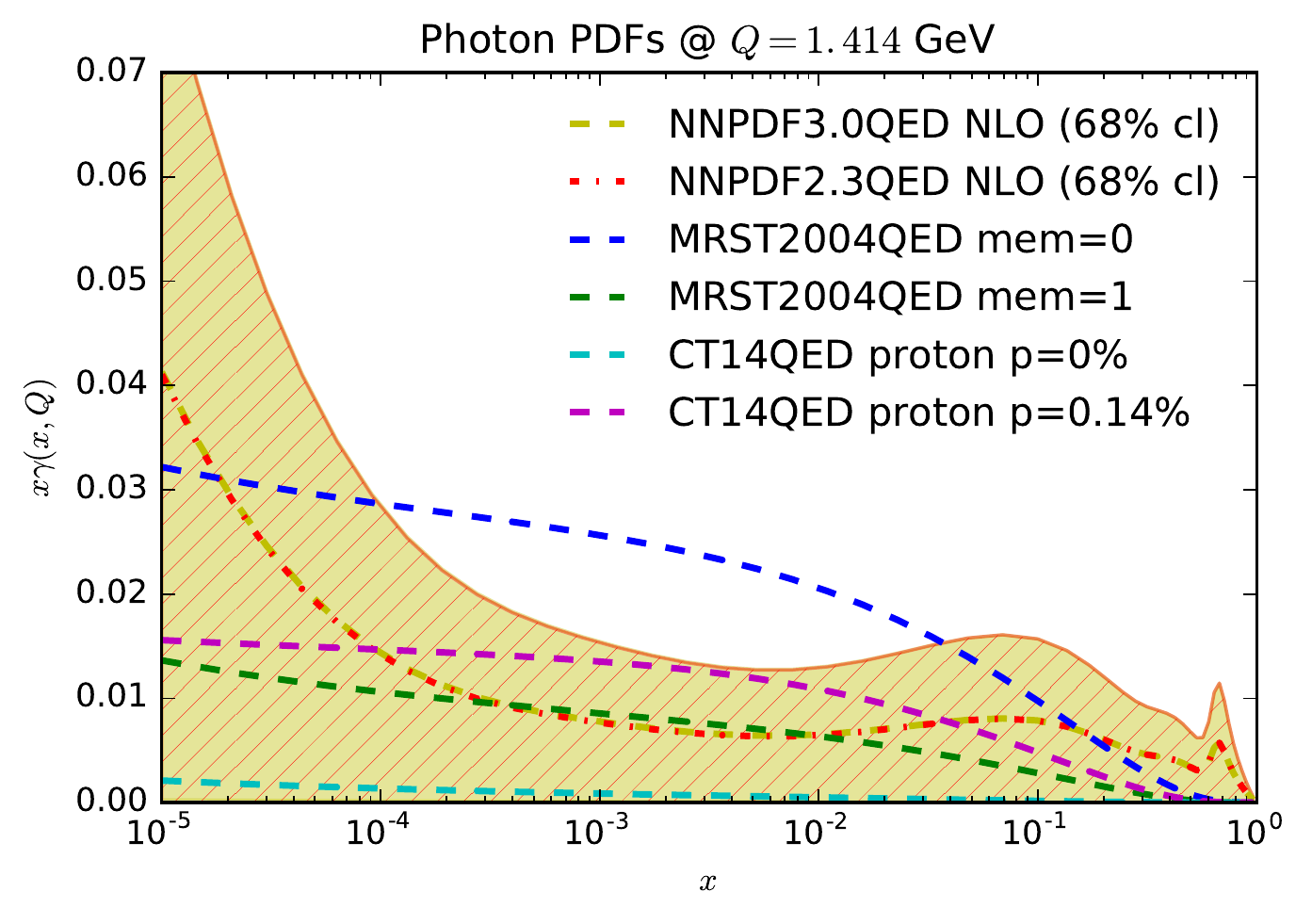}\includegraphics[scale=0.55]{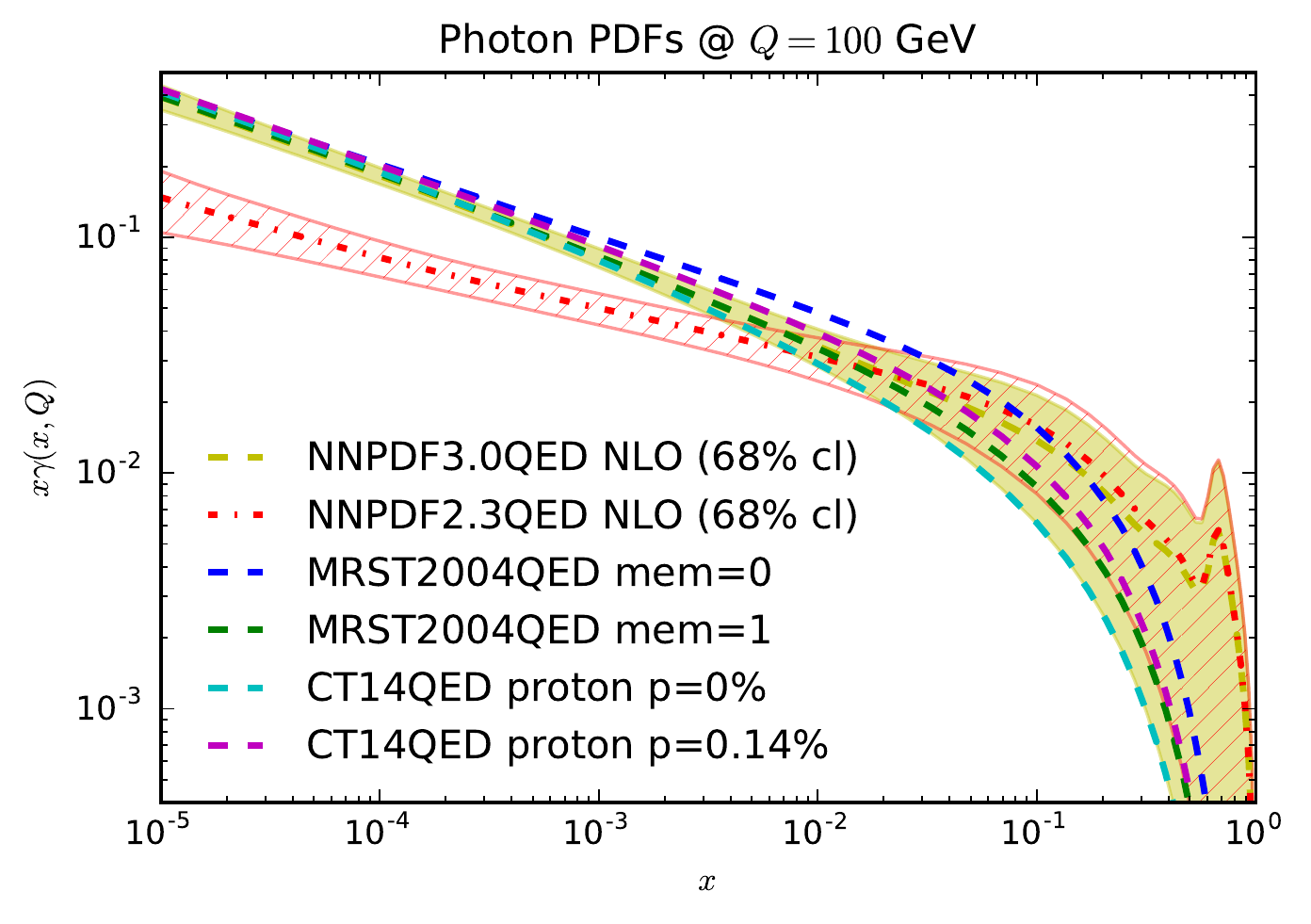}
    \par\end{centering}
  \caption{\label{fig:photon}\small Comparison of the photon PDF at
    $Q=1.414$ GeV (left) and $Q=100$ GeV (right) between NNPDF3.0QED,
    NNPDF2.3QED, MRST2004QED with ``current'' (mem=0) and
    ``constituent'' (mem=1) quark masses, and CT14QED with photon
    momentum fractions $p_0^{\gamma} = 0$\% and
    $p_0^{\gamma} = 0.14$\%. Uncertainties for NNPDF3.0QED and
    NNPDF2.3QED are given as 68\% CL level symmetric around the
    central value.}
\end{figure}

Finally, in Fig.~\ref{fig:photon} we compare the photon PDF of
NNPDF3.0QED to the other determinations currently available, namely
NNPDF2.3QED, CT14QED, and MRST2004QED. The comparison is done at NLO
because the CT14QED and MRST2004QED sets are available only at this
order. In the left panel of Fig.~\ref{fig:photon} the photon PDFs are
compared at $Q=\sqrt{2}$ GeV where, by definition, the NNPDF2.3QED and
the NNPDF3.0QED photons are identical. At this scale all
determinations are mostly compatible across the whole range in $x$
considered, with the only exception of MRST2004QED mem=1 which tends
to be a bit harder than the others for medium values of $x$. In the
right panel of Fig.~\ref{fig:photon} the photon PDFs are compared at
$Q=100$ GeV. The main observation is that, while the NNPDF2.3QED
photon PDF at small values of $x$ is substantially smaller than the
others, the photon of the NNPDF3.0QED set presents the same level of
agreement with CT14QED and MRST2004QED as at small scales. This is the
desired effect of the QCD$\otimes$QED unified evolution implemented in
{\tt APFEL}.

The extraction of the photon PDF from a complete global analysis is in
our next future plans and in this sense the NNPDF3.0QED sets are meant
to be a temporary solution. Amongst other improvements, the future
analysis will most likely include NLO QED corrections to the DGLAP
evolution equations~\cite{deFlorian:2015ujt,deFlorian:2016gvk} and new
experimental data able to constraint the photon PDF, such as the 8 TeV
high-mass Drell-Yan data from ATLAS~\cite{Aad:2016zzw}.

The NNPDF3.0QED sets are available from the NNPDF {\tt HepForge} web page:
\begin{center}
\PoSspecialurl{https://nnpdf.hepforge.org/html/nnpdf30qed/nnpdf30qed.html}
\end{center}


\paragraph{Acknowledgements}
V. B. is supported by the European Research Council Starting Grant
``PDF4BSM''. S.C. is supported by the HICCUP ERC Consolidator grant
(614577).

\end{document}